\documentclass[12pt]{article}
\usepackage{graphicx}
\usepackage[margin=1.25in]{geometry}
\usepackage[usenames,dvipsnames]{color}
\usepackage{url}
\usepackage[colorlinks = true,
            linkcolor = blue,
            urlcolor  = blue,
            citecolor = blue,
            anchorcolor = blue]{hyperref}

\newcommand\pubdate{July 2023}

\def\SLAC{SLAC,
    Stanford University, Menlo Park, CA 94025, USA}

\def\Title#1{\begin{center} {\Large #1 } \end{center}}
\def\Author#1{\begin{center}{ \sc #1} \end{center}}
\def\Address#1{\begin{center}{ \it #1} \end{center}}

\newcommand\pubblock{\rightline{\begin{tabular}{l} 
         \pubdate \end{tabular}}}
\newenvironment{Abstract}{\begin{quotation} \begin{center}
                       ABSTRACT
     \end{center}\bigskip  }{\end{quotation}}

\def\Acknowledgements{\bigskip  \bigskip \begin{center} \begin{large}
             \bf ACKNOWLEDGEMENTS \end{large}\end{center}}

\def\beq{\begin{equation}}
\def\eeq#1{\label{#1}\end{equation}}
\def\eeqn{\end{equation}}

\newenvironment{Eqnarray}%
   {\arraycolsep 0.14em\begin{eqnarray}}{\end{eqnarray}}
\def\beqa{\begin{Eqnarray}}
\def\eeqa#1{\label{#1}\end{Eqnarray}}
\def\eeqan{\end{Eqnarray}}

\def\leqn#1{(\ref{#1})}

\let\bar=\overbar

\def\lsim{\mathrel{\raise.3ex\hbox{$<$\kern-.75em\lower1ex\hbox{$\sim$}}}}
\def\gsim{\mathrel{\raise.3ex\hbox{$>$\kern-.75em\lower1ex\hbox{$\sim$}}}}

\def\del{\partial}
\def\Dslash{\not{\hbox{\kern-4pt $D$}}}
\def\dslash{\not{\hbox{\kern-2pt $\del$}}}

\def\Dlr{\mathrel{\raise1.5ex\hbox{$\leftrightarrow$\kern-1em\lower1.5ex\hbox{$D$}}}}

\def\MSB{{\bar{M \kern -2pt S}}}
\def\msb{{\bar{\scriptsize M \kern -1pt S}}}

\def\drb{{\bar{\scriptsize D \kern -1pt R}}}

\makeatletter
\def\section{\@startsection{section}{0}{\z@}{5.5ex plus .5ex minus
 1.5ex}{2.3ex plus .2ex}{\large\bf}}
\def\subsection{\@startsection{subsection}{1}{\z@}{3.5ex plus .5ex minus
 1.5ex}{1.3ex plus .2ex}{\normalsize\bf}}
\def\subsubsection{\@startsection{subsubsection}{2}{\z@}{-3.5ex plus
-1ex minus  -.2ex}{2.3ex plus .2ex}{\normalsize\sl}}

\renewcommand{\@makecaption}[2]{%
   \vskip 10pt
   \setbox\@tempboxa\hbox{\small #1: #2}
   \ifdim \wd\@tempboxa >\hsize     
       \small #1: #2\par          
     \else                        
       \hbox to\hsize{\hfil\box\@tempboxa\hfil}
   \fi}

\makeatother

\begin{document}
\begin{titlepage}
\pubblock

\vfill
\Title{Open Access: Who are the Ghost Readers?}
\bigskip

\bigskip 

\Author{Michael E. Peskin}
\medskip
\Address{\SLAC}

\vfill

\begin{Abstract}
To develop a funding model for Open Access journal publication, it is
necessary first to understand who benefits.   This is a difficult
task,because,  in Open Access, no credentials are needed to read a journal
article, and, thus, those people who access journal articles through
Open Access leave no self-identification.  We
might call these readers   ``ghost readers''.  In this paper, I propose a
method to learn the reading habits of the ghost readers.  I
explore this method using a database of downloads from the Open Access
volumes of the Annual Reviews journals. I find that the habits of the
ghost readers are 
very similar to those of academic readers from known institutions.
\end{Abstract}

\vfill

\newpage

\tableofcontents
\end{titlepage}

\newpage

\def\thefootnote{\fnsymbol{footnote}}
\setcounter{footnote}{0}

\section{Introduction}

Open Access (OA), the opening of articles in the scientific literature to anyone
who would like to read them, is obviously a benefit.  It makes journal
articles easier to obtain and offers them to a wider audience.   But,
still, OA is controversial because of the burden its puts on
publishers to find a funding model other than the traditional one of
selling their products.   The key to a successful funding model is
that those who benefit should pay the costs.  But who benefits?

The original vision of OA publishing was to make the
scientific literature accessible to new readers, readers outside
well-supported academic institutions and outside academia
altogether. For example,   the 2002 ``Budapest Open 
Access Initiative Declaration''~\cite{Budapest} begins with the philosophical
statement:  ``The public good they make possible is the world-wide
electronic distribution of the peer-reviewed journal literature and
completely free and unrestricted access to it by all scientists,
scholars, teachers, students, and other curious minds. Removing access
barriers to this literature will accelerate research, enrich
education, share the learning of the rich with the poor and the poor
with the rich, make this literature as useful as it can be, and lay
the 
foundation for uniting humanity in a common intellectual conversation
and quest for knowledge.''  To put this more bluntly, the main benefit comes to
those who are not motivated to pay, or cannot afford to. There is much
to admire in this idealistic vision, but it has not been helpful in
making OA a common practice.   Libraries are  supported to
benefit the institutions of which they are a part.   The authors of
journal articles write for their peers. These stakeholders have little motivation to
pay to bring their technical papers to readers  outside their own
institutions or networks.

But, is this perspective on the benefits of OA correct?   There is
very little evidence that has been brought to bear on this problem. 
 A well-documented
feature of OA is that the  number of downloads of OA  journal articles
increases noticeably and, sometimes,
dramatically~\cite{DavisWalters,Langham}.
However, this fact
is often not considered to be important in
and of itself. It depends on who is making these extra downloads.
In OA publication, readers meet no
paywall and so have no need to identify themselves.  We might call
these readers ``ghost readers''.  To understand who benefits from OA,
we need to learn who the ghost readers are.

One positive merit of OA might be
that it makes OA papers more prominent by increasing their citation
count.  However, it is controversial whether publishing a paper
OA increases its number of
citations~\cite{Langham,Piwowar}.  Easily obtained samples of OA
articles are
biased toward more highly cited papers, and correcting for
these biases
leaves a small effect, less than 20\% or even zero in some studies.
On the other hand, most scientific papers, and,
therefore, most citations, come from authors at leading  institutions
with well-supported libraries.  These authors can learn about and
access any papers that they wish to see, whatever the access
restrictions for others.  So it is not clear how large this advantage
could be, even in the best case.

This makes it an important problem to understand the value or lack
of value in the increase in downloads.  To do this, we must try to
identify the readers who are downloading papers using OA.  This can be
done by analyzing their reading habits.   But, often, the data needed
for such an analysis is closely held by publishers who want to
protect the details of their business models. For whatever reason,
this data is very poorly studied.

I was fortunate to be  given access to three years of data on downloads from
the journals of Annual Reviews, Inc., a nonprofit publisher, and  I
was given permission to share my analysis of it publicly.   In this paper, I
will show that, even though the new OA readers are anonymous, it is
possible to compare their reading habits to those of
well-characterized institutional readers and thus to gain evidence on
who they are.

The structure of this paper is as follows:  In Section 2, I will
explain what data on OA readership is available from the Annual
Reviews (AR) records. I will also point out some ways in which the 
AR journals are not typical scientific journals.   In Section 3, I
will present the data on the
increase in downloads for OA journals.   For AR, this has been a
pronounced effect, increasing the number of downloads by a factor of 3
or more.  I will compare this to results on downloads given in the literature.

In Section 4, I will show that the usage pattern of a body of readers
can be represented as a ``Zipf curve''.  Zipf's Law claims that
this curve is a pure power law~\cite{Zipf1932, Powers1998}.
Howver, we will see that, when applied
to journal readership, the curve has structure and provides a
fingerprint of the habits of a group of readers.  In Section 5, I will
present curves of this  types that compare the reading habits of ``ghost
readers'' to those of readers accessing AR journals through their
subscribing
institutions. It turns out that the Zipf curves for the two types of
readers are very similar.
 In Section 6, I will sharpen this conclusion by modelling the
 residual differences between the two samples.
Section 7, I will show an exception that tests the interpretation of
Zipf curves.  Section 8 will present some conclusions.

I should disclose my affliations:  I am a member of the Board of
Directors of AR.  AR is a nonprofit corporation, and the members
of the Board are not compensated for their service.  I am also 
co-Editor of the Annual Review of Nuclear and Particle Science, one of
the journals considered in this study, and I receive an annual
honorarium from AR for my work on that journal.

\section{The Annual Reviews downloads database}

\begin{table}[t]
\begin{center}
\begin{tabular}{lcc}
Annual Review of ...  &  acronym  &  OA open date \\\hline
Biomedical Engineering &   BE   &   7/13/2021         \\
Cancer Biology &   CB  &     3/9/2020         \\
Environment and Resources &  ER  &   10/19/2020           \\ 
Genomics and Human Genetics &  GG  &    9/1/2021         \\ 
Nuclear and Particle Science & NS  &  10/19/2020   \\ 
Political Science &  PS  &  5/12/2020        \\ 
Public Health &   PH  &    4/6/2017      \\ 
Virology &  VI   &       9/29/2021            \\ 
\end{tabular}
\end{center}
\caption{The eight AR journals available as OA publications in 2020-22.  These
  provide the basis for the data discussed in this paper.  The data
  from  Public Health was unfortunately taken in a different format,
  making it unusable for this analysis.   The second column shows the
date  at which the OA period began. }
\label{tab:OAjs}
\end{table}

Annual Reviews is a nonprofit publisher that produces 51 journal
volumes each year in various areas of physical, biological, and social
science. Over the past few years, AR has
conducted an experiment in OA,
making a subset of its  journals open under a plan called ``Subscribe
to Open'' (S2O)~\cite{StwoO}. In this model, if total subscription
revenue reaches a prescribed level necessary to fund the journal,
the journal is made OA for all
readers.
Using this plan,  8 of the AR journals were
converted to OA.  These journals are listed in Table~\ref{tab:OAjs}
for easy reference.   AR rolls out its journals throughout the year,
one issue per year, 
so for each of the OA journals, the starting date for OA was
different.  The data set that I will discuss includes all pdf
downloads of articles in these  OA volumes over the
three  years 2020-2022.  The main results of this paper are based on a comparison
of download records between institutional and OA readers during the OA
period.  The eight journals include journals in the physical sciences,
the social sciences, and the biological sciences.  The orientation of
these journals is in all cases quite academic, although the volumes on
Cancer Biology (CB), Genomics and Human Genetics (GG),
and Virology (VR) have some medical relevance.  Beginning in 2023, all
AR journals are being made OA through S2O.

In addition, all 51 journals were made OA during a brief period from
mid-March to mid-June 2020, at the beginning of the COVID-19
pandemic.  I will discuss some results from this period in Section~7.

The AR records refer to downloads of articles from the AR web site.
The records contain only a minimum of information:  the DOI of the
article downloaded, the date and time, the country of the reader, the reader's
institution (if the institution is a subscriber), the format
(pdf or html), and additional administrative details associated with
the download permissions.   An important property of the S2O
arrangement is that  scientists at subscribing
institutes see no difference in access between OA and non-OA
periods.  Thus, the record of downloads is populated with downloads
from identified subscribers, which can be used as a control group,
along with downloads from unidentified
OA readers.

\begin{figure}[p]
  \begin{center}
    \includegraphics[width=0.8\hsize]{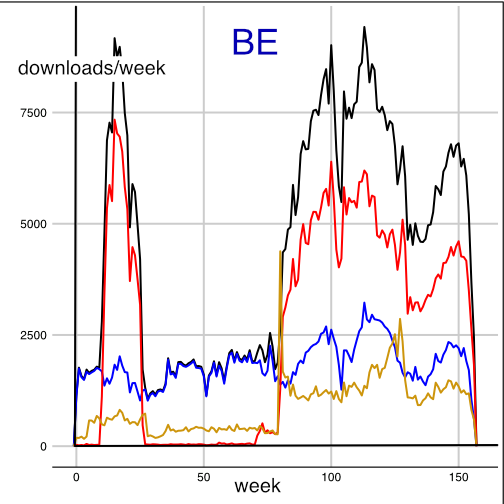}
    \end{center}
\caption{Downloads of articles from the Annual Review of Biomedical
  Engineering (BE) per week for the period 2020-2022.   The three
  colored curves correspond to the following classes of readers:  blue
  - readers from identified subscribing institutions; red - Open
  Access readers (Ghost readers) without institutional identification;
  gold - readers of specific articles made ``Free to Read'' by Annual
  Reviews. The black curve is the sum of Institutional and Ghost
  readers.
  The red downloads turn on in two different
  periods---during the first three months of the pandemic and
  during a later period in which the journal was made Open Access
  through Subscribe to Open ~\cite{StwoO}.}
\label{fig:BEtimeline}
\end{figure}

A typical usage curve is shown in Fig.~\ref{fig:BEtimeline}. 
The blue, red, and gold curves show the downloads  per week from  from
AR
Biomedical
Engineering (BE) for three different
classes of readers, which I will refer to as ``Inst'', ``Ghost'', and
``Free'':
\begin{itemize}
\item Inst readers (blue):   Readers from subscribing institutions.
\item Ghost readers (red):  OA readers, identified only by country of origin.
\item Free readers (gold):   Readers of ``Free Access'' papers.
\end{itemize}
  The
black curve gives the sum of the downloads per week from Inst and
Ghost readers.
This association of these three types of readers with colors is used
consistently in the figures below.  Inst readers are those whose
downloads are registered with a
subscribing institution.   Ghost readers are those who are granted
access but are not identified with an institution; these are the Open
Access readers.   Free readers are those who download papers declared
``Free Access''  by AR. This designation is used by the AR marketing department
to highlight especially topical papers in a given journal in advance of the annual
journal release, and to make available articles that are discussed in
widely read sources such as the New York Times.   In some developing
countries, entire volumes are made Free Access.  In general, though, these
Free Access papers are mainly articles that are expected to be
especially interesting to the general public.

AR Public Heath, the first journal that AR made OA, was actually made
free to access in 2017, before the start
of S2O.  Unfortunately, even after it became an S2O journal, all of its
articles were classified as ``Free Access'', making its download record
unusable for the analysis in this paper.

The data in  Fig.~\ref{fig:BEtimeline} shows the increase in journal usage during
the pandemic OA period, from March to June 2020 (weeks 11 to 24), and
the steady increase in journal usage during the period
of OA under S2O (week 81 onward).  The usage of the other journals
shows a similar pattern, as I will discuss in the next section.  The
data set used for analysis in subsequent sections
is the list of all downloads from the beginning
of the S2O period to the end of December 2022.  For
two of the journals (CB and PS), the S2O period began before or during
the pandemic OA period.

\section{The OA Effect on Downloads}

A remarkable feature of Fig.~\ref{fig:BEtimeline} is the tremendous
increase in the number of downloads from the addition of Ghost
readers.  Their usage increases the total usage by about a factor of
2.5.  Notice that the blue curve representing
the subscribers  is approximately constant through OA
and paywalled periods.  Thus,  OA access does not cannibalize the
institutional  usage but simply adds to it.

\begin{figure}
\begin{center}
 \includegraphics[width=0.4\hsize]{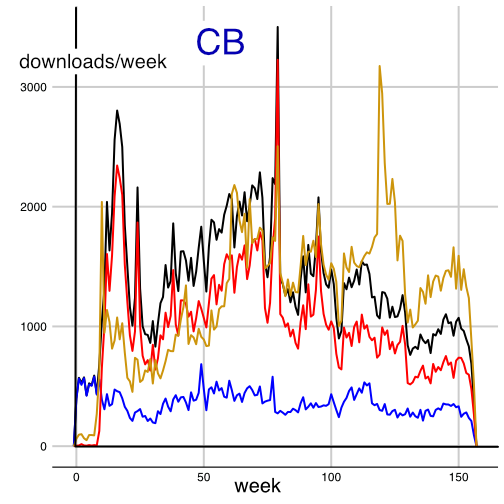} \ \ 
 \includegraphics[width=0.4\hsize]{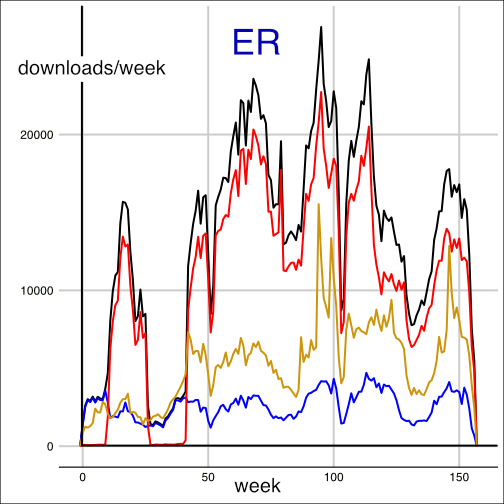} \\
 \includegraphics[width=0.4\hsize]{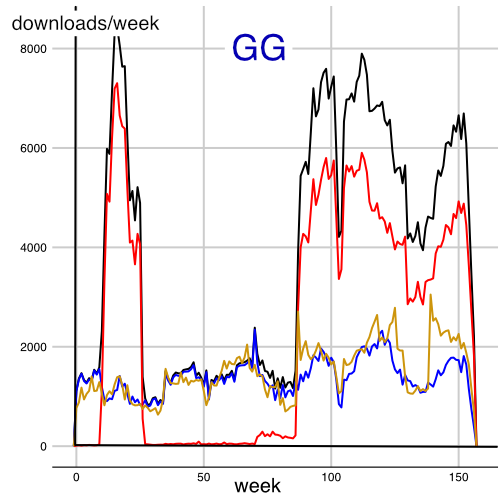} \ \
 \includegraphics[width=0.4\hsize]{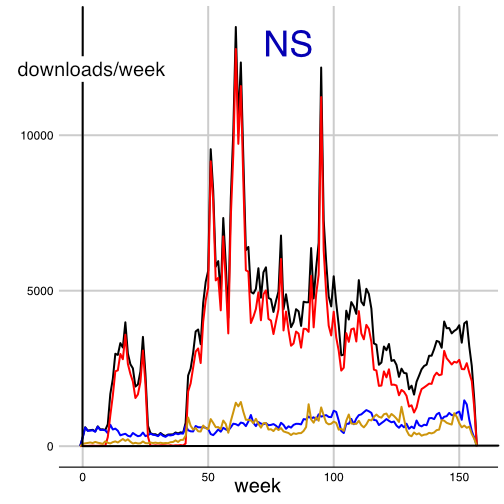}\\
 \includegraphics[width=0.4\hsize]{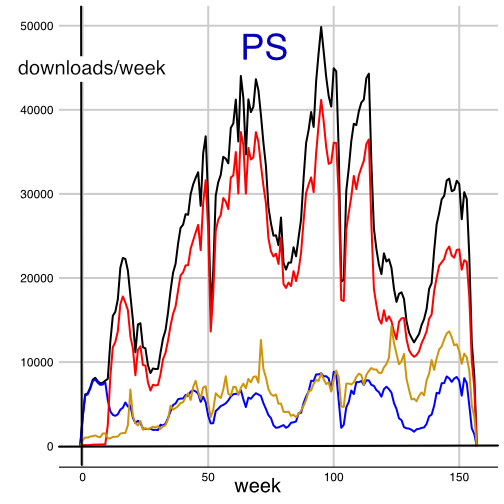} \ \
 \includegraphics[width=0.4\hsize]{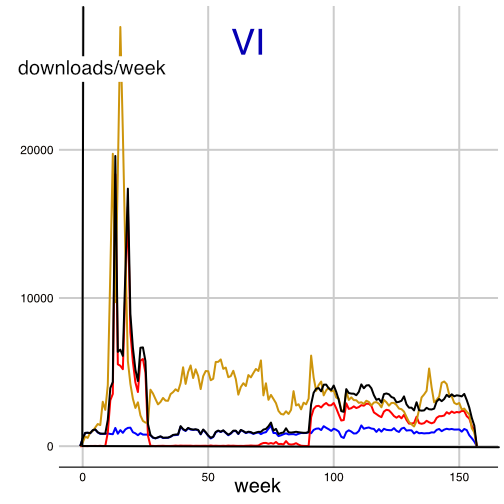}
\end{center}
\caption{Downloads per week of articles from the other 6 Annual Reviews
  journals made Open Access through S2O.
  The notation is as in Fig.~\ref{fig:BEtimeline}.}
\label{fig:timelines}
\end{figure}

This general behavior is typical
for the OA journals.  In Fig.~\ref{fig:timelines} I show the
corresponding figures for the other six S2O journals.   For each, there
is a different start date for the OA period, but, after than date, the
increase in the  level of downloads proceeds in a consistent way.

\begin{figure}[p]
\begin{center}
 \includegraphics[width=0.9\hsize]{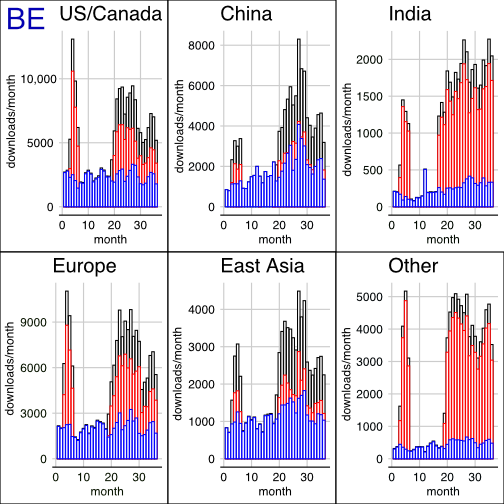} \ \ 
 \end{center}
\caption{Downloads per month  of articles from the Annual Review of Biomedical
  Engineering (BE) during the S2O period in  2020-2022, for Inst and
  Ghost readers, for six
  different regions of the world, as described in the text.  The
  colors are as described in  Fig.~\ref{fig:BEtimeline}. The histograms
  are not stacked; rather, the black histogram shows the sum of Inst
  and Ghost downloads.}
\label{fig:BEglobal}
\end{figure}

\begin{table}[t]
\begin{center}
\begin{tabular}{lccc}
AR journal  & Total/Inst - global  & Total/Inst - US/Canada
  & Total/Inst - India \\ \hline
  BE   &     2.47          &     2.30      &
  4.89\\
  CB                &       3.78          &  3.34
                       &  7.57 \\
 ER  &    5.13         &  4.37      &
                                                                  17.74 \\ 
 GG  &  2.80      &    2.66     & 7.23 \\ 
NS  &  5.44       &     4.69      &
                                               6.24\\ 
  PS     &    4.99           &   4.35       &    37.28         \\ 
VI            &       2.72
  &   2.43    &  13.05  \\ 
\end{tabular}
\end{center}
\caption{For each AR journal, the ratio of total  downloads (Inst + Ghost
readers) to downloads from Inst readers only during the S2O period in
2020-22.  The first column is ratio of the totals for all regions.
The second and third columns give the ratios for US/Canada and for India.}
\label{tab:OArats}
\end{table}

The readership of the AR journals is global, so it is interesting to
see the global distribution of downloads.
Fig.~\ref{fig:BEglobal} 
shows the breakdown of the timeline (by month) for BE for six groupings of
countries: US and Canada, Europe, China, India, East Asia (excluding
China), and all others.  The figure compares the usage
from Inst and Ghost readers for each grouping.  The largest relative
increase occurs in India and in the Other category that includes most of
the developing world.  There, the increase in downloads is close to a
factor of 5.  This accords with the idealistic mission  of
OA.  However, the increase in readership in the US and Canada and in
Europe is very substantial and actually dominates in terms of absolute
numbers of downloads.   The increase here of a factor of more than 2
is of 
direct relevance to libraries at North American academic
institutions. 

In Appendix A, I present the corresponding comparisons for the other
six OA journals.  The ratios of total to Inst downloads is recorded in
Table~\ref{tab:OArats}.  Though the ratio of Ghost to Inst readers varies
among these journals, the general pattern holds in all cases.   In
particular,  the global ratio is close to the North American ratio,
indicating the numerical dominance of downloads from developed regions.

 I note
that the increase in readership due to OA shown here is larger than
that reported in other studies.  The studies \cite{Nicholas,Davis2,Davis2011}
report increases in downloads of about a factor 2.  In
\cite{DavisWalters}, Davis and Walters are concerned about the effect
of robots  on the download counts.  In their study \cite{Davis2}, they
remove downloads from identified robots, and this decreases the effect
by a factor of 2 (though it is still an 80\% increase).  They write
about \cite{Nicholas} that ``most of this increase was attributed to
Internet robots (automated applications that index web pages) rather
than human intention''.  However, in the paper itself, Nicholas and
colleagues conclude ``There has been a substantial shift in the make-up
of users by referrer link - most notably there has been an increase
in the share of accesses via a search engine.... This represents a
dramatic
and major shift in scholarly information seeking behaviour.''  This is a very
different statement.  Today, journal web sites can prohibit
indiscriminate downloads by robots.   I will show in Section 6 that
robots have at most a very small influence on the AR data.

The situation of AR is special in
some respects.  AR does not publish primary literature but rather presents
reviews covering areas of science.  The study \cite{Davis2} found a
positive
correlation between review articles vs. primary articles and the
number
of downloads.  Reviews are often the
starting point for students or established researchers who seek to
learn about or enter a new  specialty. They are a source of references
to primary research articles, so some of the usage might be
simply scanning
for a pointer to a reference.   The  AR journals are individualy
well-known,
typically
among the top 5 journals in their respective areas in terms of impact
factor; see \cite{impactfactor}.   It would be very interesting to
analyze data of the sort that I discuss here over a wider range of journals.

Still, I hope that those who view this data will think more seriously about
the simple fact of the increase in usage.  If someone offered you, as
a librarian, a magic trick that, at minimal cost, would double or
triple
the usage of a journal that you are supporting,
wouldn't you grab for it?

\section{Zipf curve analysis}

To understand better the identities of the Ghost readers and their
relevance for journal usage, we can look into their behavior. 
Despite the fact that we know nothing about the Ghost readers except
which articles they download, we can still make use of this minimal
information to learn about them. 

Zipf's law~\cite{Zipf1932,Powers1998} is the statement that, in a list of words in a
text or other items of relevance in a corpus of data, the frequency of
appearance decreases as a power of the rank.   This is, of course, an
idealization.  In a real data set, the frequency vs. rank distribution can
have structure.  In that case, the distribution of frequency vs. rank, on
a log-log plot, can give information on what partitions exist in the
dataset according to the usage it describes.   In the rest of this
paper, I will refer to the log-log plot of frequency vs. rank as the
``Zipf curve''.

\begin{figure}
\begin{center}
 \includegraphics[width=0.5\hsize]{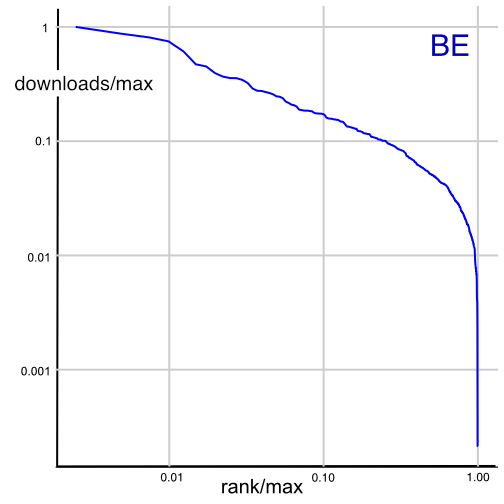} \ \ 
 \end{center}
\caption{Zipf curve, the plot of downloads vs. rank, for Inst readers
  of the AR Biomedical Engineering (BE) journal.}
\label{fig:sample}
\end{figure}
\begin{figure}
\begin{center}
 \includegraphics[width=0.5\hsize]{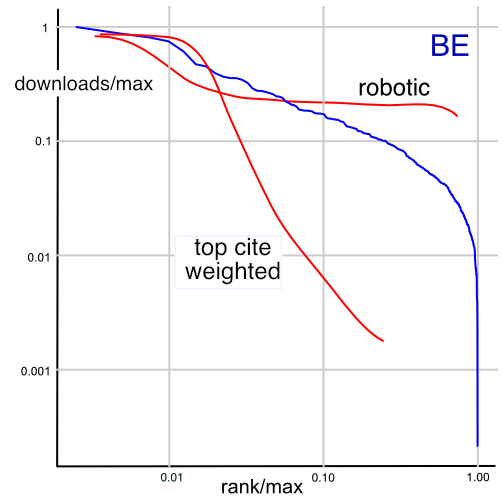} \ \ 
 \end{center}
\caption{Zipf curve, the plot of downloads vs. rank, for Inst readers
  of the AR Biomedical Engineering (BE) journal, plus some conjectures
  for the form of the Zipf curve for Ghost readers.}
\label{fig:sampleplus}
\end{figure}

It is interesting to make this plot for the downloads data.  In
Fig.~\ref{fig:sample}, I show a  plot of the number of copies of
each paper in the BE  journal
selected by Inst readers. The download numbers and ranks are normalized so that
the largest number of downloads and the largest rank are rescaled to 1.
Fig.~\ref{fig:sampleplus} adds some conjectured curves for the Ghost
readers.  If the OA readers are mainly interested in the most highly
used or 
cited papers, their curve would fall off more sharply.   If the OA
readers are blindly searching for information (or are bots rather than
humans), their curve would be more flat.

\begin{figure}
\begin{center}
 \includegraphics[width=0.7\hsize]{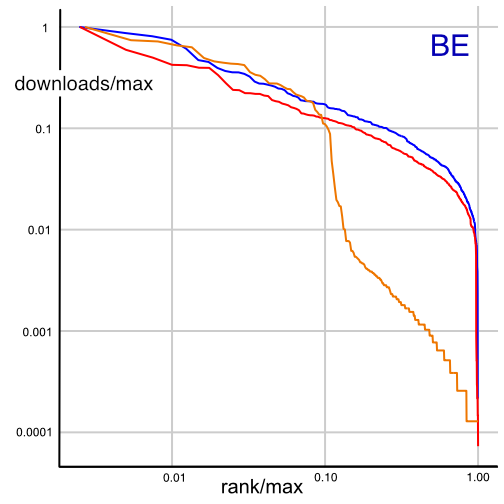} \ \ 
 \end{center}
\caption{The actual Zipf curve, the plot of downloads vs. rank, for readers
  of the AR Biomedical Engineering (BE) journal.  The color coding for
  Inst, Ghost, and Free readers is the same as that in
  Fig.~\ref{fig:BEtimeline}, that is blue for Inst readers, red for
  Ghost readers, and gold for Free readers.}
\label{fig:BEusage}
\end{figure}

The actual curves for BE for the three types of readers
are shown in Fig.~\ref{fig:BEusage}. Notice
that the gold curve, for Free Access readers, does show the expected
behavior of falling off rapidly with rank.  However, the curve for
Ghost readers closely follows the curve for Institutional
readers. The individual data sets for this and other journals
contain hundreds of thousands of of downloads, millions in some cases, so
the statistical errors on the curves are small compared to the
differences between the Inst and Ghost lines.

\section{Results from the Zipf curves}

\begin{figure}
\begin{center}
\begin{center}
 \includegraphics[width=0.4\hsize]{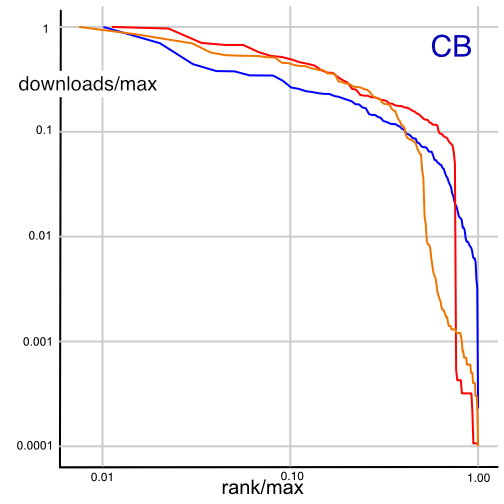} \ \ 
 \includegraphics[width=0.4\hsize]{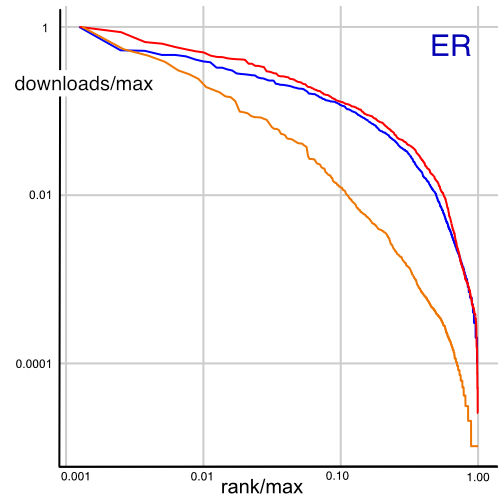} \\
 \includegraphics[width=0.4\hsize]{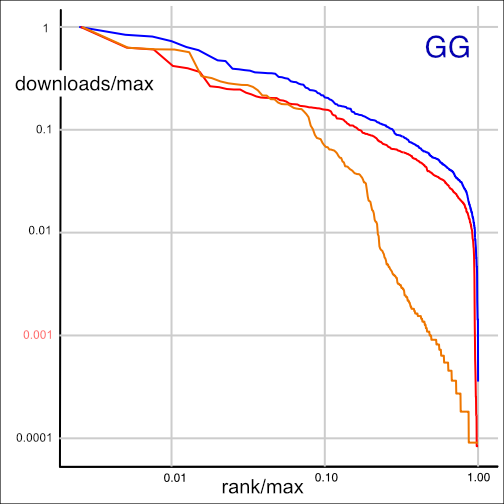} \ \
 \includegraphics[width=0.4\hsize]{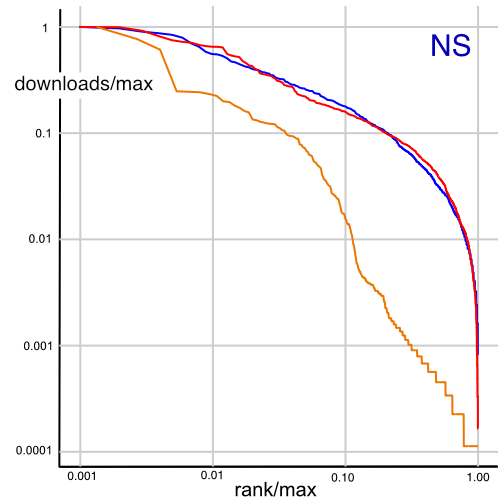}\\
 \includegraphics[width=0.4\hsize]{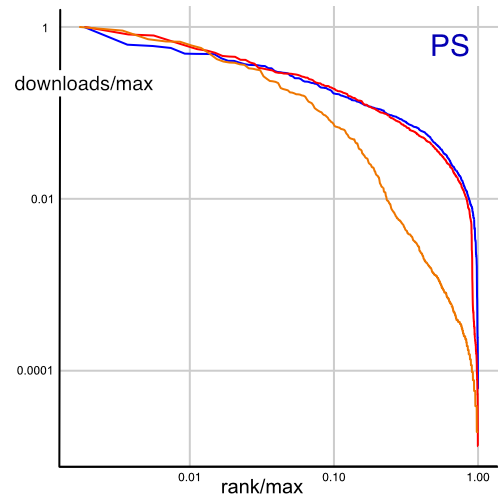} \ \
 \includegraphics[width=0.4\hsize]{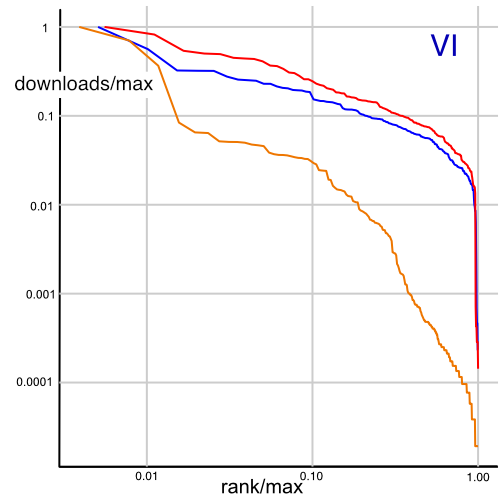}
\end{center}
 \end{center}
\caption{Zipf curves for the other 6 Annual Reviews
  journals made Open Access through S2O.  The color coding  is as in
  Fig.~\ref{fig:BEusage}.}
\label{fig:GIsix}
\end{figure}

In Fig.~\ref{fig:GIsix}, I show the comparisons of the Zipf
curves against total downloads for the other six OA journals.  In all
of the plots, the curves for Ghost readers are very similar to the
curves for Inst readers. For ER, NS and PS, the curves are practically
indistinguishable. The rank for each paper is the rank within
that
class, so the same paper might have a different rank for each of the
three types of readers.  However, we will see in Section 7 that the
ranks for Inst and Ghost readers are highly correlated.

While there is a tendency for the Ghost curves for the
medically relevant journals CB and VI to have a lower slope and thus
lie above the Inst curves, 
the effect is not pronounced, and
the distribution is still far from random selection.  I will
investigate in Section~7 to what extent this is a meaningful effect.

\begin{figure}
\begin{center}
 \includegraphics[width=0.9\hsize]{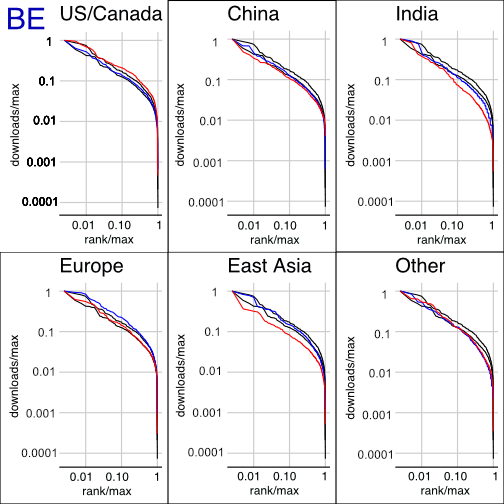} \ \ 
 \end{center}
\caption{Zipf curves for the AR Biomedical Engineering (BE) journal,
  shown separately for each of the six regions in
  Fig.~\ref{fig:BEglobal}.
  The color coding  is as in Fig.~\ref{fig:BEusage}.  The black curves are
  the two Zipf curves for the total sample shown in that figure.}
\label{fig:BEuregions}
\end{figure}

It is possible to get a better feeling for the variation in these
plots by comparing the results for different regions.
Fig.~\ref{fig:BEuregions} shows the separate usage plots for the six
regions defined in Fig.~\ref{fig:BEglobal}.  In these plots, the
black curves are the Zipf curves for the total sample, from
Fig.~\ref{fig:BEusage}, while the
blue and red curves show the curves for the Inst and Ghost readers in
that region.   Note that the ordering of the red and blue curves is
different in the US/Canada data than in other regions.
The variation among the six regions in these plots is as
large as the difference between the Inst and Ghost curves for the full
sample. This feature is also seen for the other five journals.  The
plots for the other 6 journals are given in Appendix A.

\section{Modelling of the Zipf curves}

Now that we have seen that the Zipf curves for the Ghost and Inst
readers are quite similar, it is interesting to probe further for the
cause of their small differences.   The journals of medical interest, CB
and VI, show Ghost reader curves that deviate in the positive
direction from the Inst reader curves.  Is this a reflection of new
readers interested in the results of medical relevance, or is this a
spurious effect that does not reflect the nature of the field?  We can
investigate this by trying to make a model of the relation between the
Inst and Ghost curves.

\begin{figure}
\begin{center}
 \includegraphics[width=0.8\hsize]{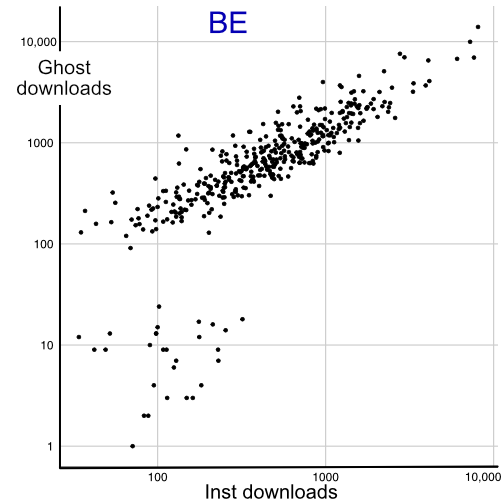}
 \end{center}
\caption{Scatter plot of the number of Inst downloads vs. Ghost downloads for AR
  Biomedical Engineering (BE). Each point corresponds to one paper.}
\label{fig:BEscatter}
\end{figure}

A useful first step is to examine the scatter plot of Inst downloads
vs. Ghost downloads paper by paper.  Do the Inst readers and Ghost
readers prefer the same papers?   For BE, the comparison is
shown in Fig.~\ref{fig:BEscatter}.  For this and the other journals,
only a handful of papers are downloaded only by one group; these
are omitted from the analysis. The plot shows a strong
correlation between the papers chosen by the two sets of readers.
The cluster of papers at very small download numbers turns out to be
the set of papers in the most recent two annual volumes.  Articles
from the 2022 volume obviously could not be downloaded in 2020-21.
More generally, it takes some time for the search engines to
catch up to the existence of the most recent articles.  In performing fits to
this and similar distributions, I disregard these low-download points
and use the data in the top cluster
only.

Looking more closely, the ratio of the number of downloads between
the highest- and lowest-cited papers in the major group is larger for
the Inst readers than for the Ghost readers, a factor of about $10^4$
for the Ghost downloads and about $10^5$ for the Inst downloads.
The effect is well
described by a simple linear fit of the logarithms of the downloads,
corresponding to
\beq
GD =  A (ID)^\alpha
\eeq{GDID}
where $ID$ and $GD$ are the downloads of a given paper from the Inst
and Ghost readers, respectively.   The same pattern is seen in the
other 6 journals.  Table~\ref{tab:alpha} gives the fitted values of
the exponent $\alpha$ for the 7 S2O journals. All fits here and below
are done
using the {\tt lm} package
in R~\cite{lm}. Fitting to a higher-order polynomial gives a minor
improvement of the fit,
not enough to overcome the clarity of the interpretation in
\leqn{GDID}. The ranks of downloaded papers  are also
highly correlated.  The last column of the table gives the Spearman
rank correlation coefficient~\cite{Spearman} for each data set.  By
this measure, the Inst and Ghost readers rank the downloaded papers in
almost the same way.

In all cases, $\alpha$
is smaller than but close to 1.   It is remarkable that the values of
$\alpha$ do not depend on the nature of the field in any systematic way.
The scatter plots for the other 6 journals are given
in Appendix A.

\begin{table}[t]
\begin{center}
\begin{tabular}{lcc}
AR journal  & $\alpha$  &  $\rho$
                                                   \\\hline
 BE   &  $   0. 814 \pm 0.021$ &   $   0.88    $       \\
  CB  &   $      0.832 \pm 0.050$   & $ 0.90 $     \\
  ER  &     $      0.798 \pm 0.014$   & $ 0.93 $     \\
  GG  &     $      0.788 \pm 0.022$   & $ 0.87 $   \\
NS  &       $      0.787\pm 0.015$   & $ 0.88  $   \\
 PS  &      $      0.855 \pm 0.025$   & $ 0.84  $  \\
  VI   &    $      0.910 \pm 0.034 $  &  $0.90 $   \\
   \end{tabular}
\end{center}
\caption{Best fit parameter $\alpha$ for the 7 S2O journals, corresponding to the
  equation \leqn{GDID} in the text, and the Spearman rank correlation
  $\rho$ for each data set. The parameter $A$
  in \leqn{GDID} has no effect on the Zipf curve.} 
\label{tab:alpha}
\end{table}

\begin{figure}
\begin{center}
 \includegraphics[width=0.7\hsize]{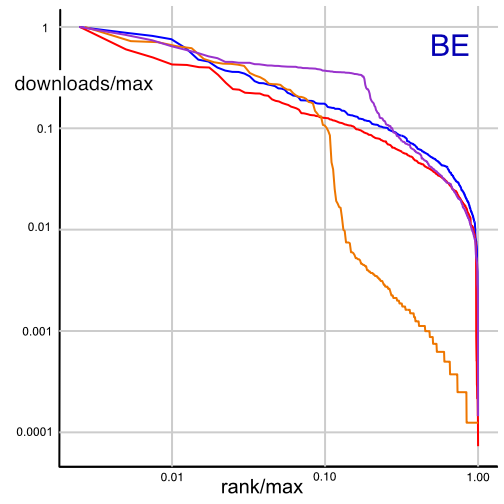} \ \ 
 \end{center}
\caption{Unsuccessful attempt to model the Zipf curve for Ghost
  downloads from the BE journal using a robot randomly sampling the
  most recent 4 volumes.  The purple line is the model.}
\label{fig:robot}
\end{figure}

Notice that there is no evidence for robotic downloads in the Ghost
reader data.   A robot that downloads indiscriminately from a subset
of the journal issues would leave an imprint as a horizontal band on
the scatter plot, with many articles at roughly the same number of
Ghost downloads independent of the number of downloads from the presumably
better informed Inst readers.  An important feature that checks this
is that the Ghost readers disfavor the same papers as the Inst
readers.  An indiscriminate robot would randomly pick up some of these
disfavored papers.  My attempts to fit the Zipf curves with a fraction
of the Ghost downloads from robotic downloads from a subset of the
volumes give results of the form shown in Fig.~\ref{fig:robot}.  This is not the
right way to obtain a good fit to the Zipf curve data.

Because the Ghost readers have a smaller variation in downloads than
the Inst readers, one might expect that the Zipf curves for the Ghost
readers to remain higher for most of
the range in $\log$(rank), so that the 
Ghost curves (red) would typically lie above the Inst curves (blue).    Looking
over the results in Figs.~\ref{fig:BEusage} and \ref{fig:GIsix}, it
is clear that this is not always the case.  I have tried quite hard to
find a causal mechanism for the differences without finding a
compelling answer.  However, there is an important  source of random
variation among the
curves.  A difficulty with using the ratio rank/maximu rank  as a
variable
is that  the number of
downloads
for a given rank can depend strongly
on the  number of downloads for the few most-downloaded
papers. It can be seen from the scatter plots that the ratio of Ghost to Inst
downloads of  the most downloaded papers often vary substantially from
the fit to eq. \leqn{GDID}, and sometimes these leading papers are not
even the same in the two samples.  This suggests that the overall relative
normalization of the Inst and Ghost curves can fluctuate in a random way.

\begin{figure}
\begin{center}
 \includegraphics[width=0.8\hsize]{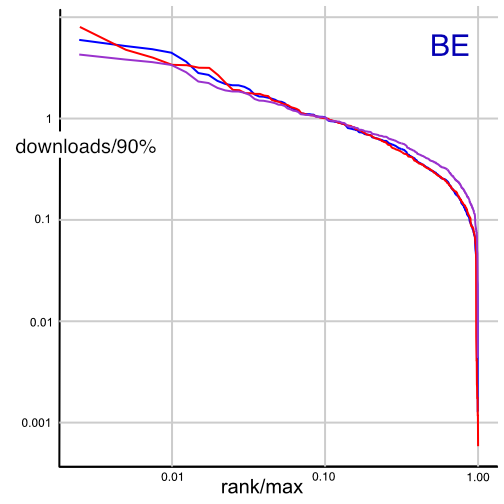} \ \ 
 \end{center}
\caption{Fit to the Zipf curve for Ghost
  downloads from the BE journal normalizing downloads to the value at
  the 90th percentile in each distribution.   The blue curve is simply
  a shift of the original Zipf curve for the Inst readers.  The violet
  curve transforms the Inst curve using the formula  \leqn{GDID}
  to convert Inst downloads to
  Ghost downloads.}
\label{fig:BEfit}
\end{figure}

A way to test this is to normalize the number of downloads to a paper
that is deeper in the downloads distributions.   To avoid picking a
paper from the very large number with few downloads, I have chosen to
normalize to the number of downloads at 90th percentile in each
distribution. This number is typically about 20\% of the maximum
value.  (Normalizing at the 75th percentile gives similar
curves.)   I show
the results for
  BE in Fig.~\ref{fig:BEfit}.  The change in
  reference value essentially removes the discrepancy between the Inst
  and Ghost curves.   In the figure, the blue line is the original
  Inst curve simply rescaled.  The purple curve is the tranformation
  of the Inst curve using the formula \leqn{GDID} to convert Inst
  downloads to Ghost
  downloads.  This gives a small improvement in the description of the
  Ghost curve (red) over part of its range.  The similar curves for the other 6 journals
  are given in Appendix A.  In general, it is not clear whether the
  blue or purple curve is a better model for the red curve.
  The two hypothesis are quite similar due to the
  closeness of $\alpha$ to 1. Also, the slope of the relation between
  Inst and Ghost downloads does depend on the rank in a way that is
  not captured by the simple relation \leqn{GDID}.  Whichever model is chosen,
  the similarity of the fitted curves to the true Ghost download data
  shows that the differences seen in Section 5 come almost entirely 
 from the fluctuations in downloads of the very  top-cited papers. They
 are not a real effect rather just reflect the luck of the draw.
 
  There are differences in behavior between the Inst and Ghost
  readers.  One indication of this is the narrower range of download
  values characterized by $\alpha < 1$.   Also, as noted already in
  \cite{Davis2},  the Ghost downloads contain a significantly higher
  fraction of html vs. pdf than the Inst downloads, possibly
  indicating a tendency for immediate reading or checking references.
  These tendencies  might indicate different readers, but they might
  also indicate the same readers carrying out different tasks at
  different times of the day.  Even with these small differences, one
  cannot escape the striking similarities of the Inst and Ghost samples.

\section{Pandemic-period results}

\begin{figure}
  \begin{center}
    \includegraphics[width=0.4\hsize]{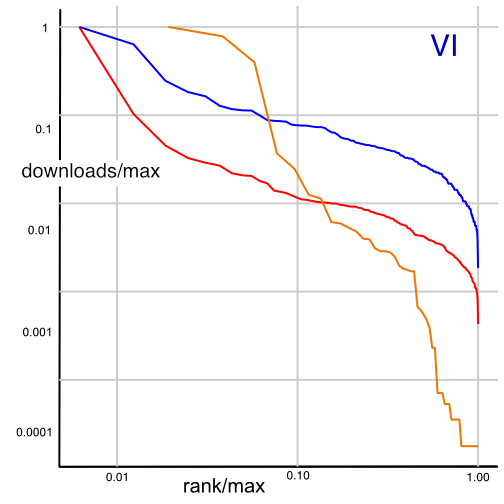} \ \ 
 \includegraphics[width=0.4\hsize]{figs/VIusage.png} 
 \end{center}
\caption{Zipf curves for the AR Virology (VI) journal, for the period
  of Open Access during the pandemic (left) and the period of S2O Open
  Access (right). 
  The color coding  is as in Fig.~\ref{fig:BEusage}.}
\label{fig:Corona}
\end{figure}

\begin{figure}
\begin{center}
    \includegraphics[width=0.4\hsize]{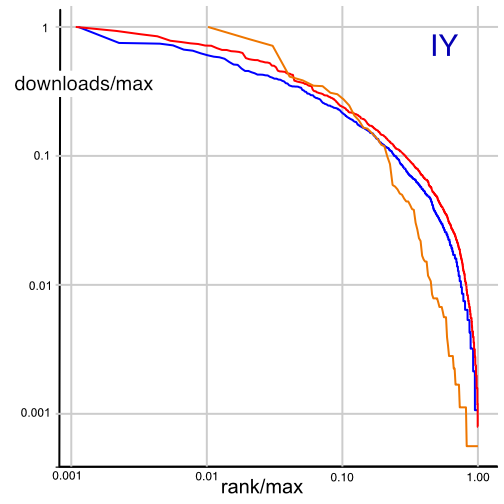} \ \ 
 \includegraphics[width=0.4\hsize]{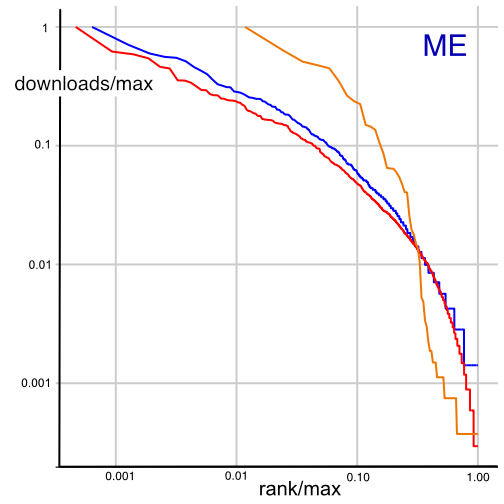} 

 \end{center}
\caption{Zipf curves for the AR Immunology journal (left) and the AR
  Medicine journal (right), for the period
  of Open Access during the pandemic.
  The color coding  is as in Fig.~\ref{fig:BEusage}.}
\label{fig:ImmMed}
\end{figure}

It should be asked whether the qualitative similarity of the Inst and
Ghost Zipf curves is ever broken in the AR data.  There is one
interesting example of this.   During the early months of the pandemic,
AR  made many of its articles on coronaviruses free to read (Free
Access).  This may have driven readers to explore other articles in AR
Virology. As I noted in Section 2, all of
the AR journals were made OA for a period of 3 months from mid-March
to mid-June 2020.  One might then expect that the Ghost curve for VI
in this period would be highly skewed toward papers on viral diseases.
Fig~\ref{fig:Corona} compares the Zipf curves for VI in the pandemic 
OA period and the period of S2O access beginning in November 2021.
The difference is striking, with the pandemic downloads clearly
showing concentration on the most highly downloaded articles.  Oddly,
this effect is not seen in the usage of AR Immunology (IY) and AR
Medicine (ME)
during the pandemic period, as shown in Fig.~\ref{fig:ImmMed}.  Both
journals do show a high usage of Free Access articles.

\section{Conclusions}

In this paper, I have examined the reading habits of the Open Acess
``Ghost readers'', readers who access OA papers without leaving
identifiable signals of their identity.  I find that, for usage of the
AR journals Open Access under the S2O policy in 2020-2022, the
behavior of the Ghost readers is very similar to that of readers from
identified academic institutions.  This is true over the 7 journals
from diverse fields for which data is available.   The same pattern is
seen for readers from all regions of the world, both in developed and
emerging nations.

The simplest explanation for the evidence presented here is that the 
``Ghost readers'' are not ghosts at all.  They are educated scientists, our
colleagues and students.  Open Access simply makes it easier to 
access the articles that the scientists in this group need to do
their research.

There are other indications that working scientists will access papers
more frequently when paywalls are absent.  The most striking example
of this is the large usage from North America of the illegal pirate
site Sci-Hub, as reported by John Bohannon in
Science~\cite{SciHub}. In this
article, Bohannon points out not only that a large fraction of Sci-Hub
usage comes from the US, but also that this usage is concentrated near
US academic centers.   I do not find these results surprising.  As a
working scientist, I know that when I am hot on the trail of an idea
and I see a reference to a paper that has relevant information, I need
to have it.  If I am not in my office and
the paper is blocked by a paywall, I am quick to seek
a publicly available version.  In my field of high-energy physics,
almost all journal articles are available in preprint form on the
Cornell arXiv~\cite{arXiv}.  Other scientists are not so fortunate.
It would be a great benefit to publishers, libraries, and researchers if the easiest way
to access any journal article would be from the journal website itself.

If indeed the ``Ghost readers'' are us and our colleagues, we must
take seriously the increase in usage that is a well-documented feature of
Open Access publication.  If Open Access increases the utility of journal articles, it
is a positive benefit to all readers, but, in particular, to
university scientists whose libraries are responsible for the support
of their journals.  I have shown that, for the case of the AR
journals, this benefit is strongly felt not only for those readers who
cannot afford access but also for scientists in regions where their
work is strongly supported by government and by universities.  It
would be very illuminating to carry out studies of the sort that I
have presented here for other types of journals and, in particular,
for journals that present primary research articles. 

The S2O model of Open Access favored by Annual Reviews continues to
put the 
burden of paying for journals on university librarians.   But, it does
not increase these costs, and it offers increased local usage as a
benefit.  It also gives librarians agency to decide which journals
they will support to keep in business with their limited funding.

 To the extent
that the conclusions of this paper  hold broadly for scientific
journals, they make a strong case for Open
Access publication that is independent of the idealistic case
described at the beginning of this paper. If  Open Access benefits all of
us in the academic community, and if it can come as a benefit with no
increase in expense, we academics should make it our standard
practice.

\Acknowledgements

I am grateful to Paul Calvi, Oren Bergman, Roy Marucut and Richard
Gallagher at Annual Reviews for making the database described in this
paper available to me and for assistance with its use.   I thank Paul
Ginsparg, Kamran Naim, and Laura
Peskin for advice and comments on the manuscript.
The data analysis in this paper was done using Hadley Wickham's
``tidyverse''~\cite{tidyverse}.

\newpage

\appendix

\section{Additional figures}

Here are the supplementary figures referred to in the text.

\begin{figure}[b]
\begin{center}
 \includegraphics[width=0.9\hsize]{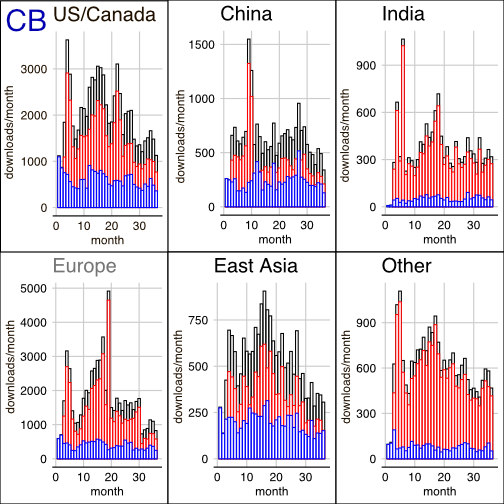} \ \ 
 \end{center}
\caption{Downloads of articles from the Annual Review of Cancer
  Biology (CB) per month for the period 2020-2022, for Inst and
  Ghost readers, for six
  different regions of the world, as described in the text.  The
  colors are as described in  Fig.~\ref{fig:BEglobal}. The histograms
  are not stacked; rather, the black histogram shows the sum of Inst
  and Ghost downloads.}
\label{fig:CBglobal}
\end{figure}

\begin{figure}[p]
\begin{center}
 \includegraphics[width=0.9\hsize]{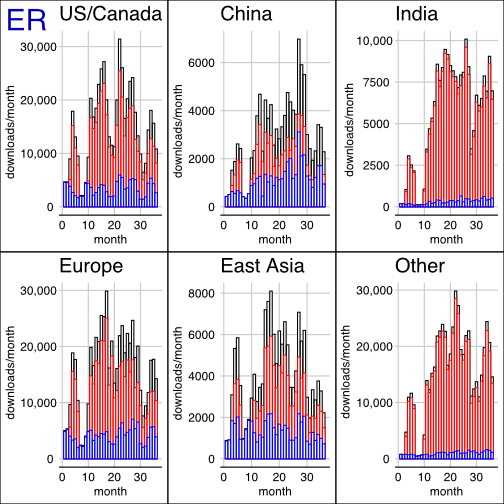} \ \ 
 \end{center}
\caption{Downloads of articles from the Annual Review of Environment
  and Resources  per month for the period 2020-2022, for Inst and
  Ghost readers, for six
  different regions of the world, as described in the text.  The
  notation is as in Fig.~\ref{fig:CBglobal}.}
\label{fig:ERglobal}
\end{figure}

\begin{figure}[p]
\begin{center}
 \includegraphics[width=0.9\hsize]{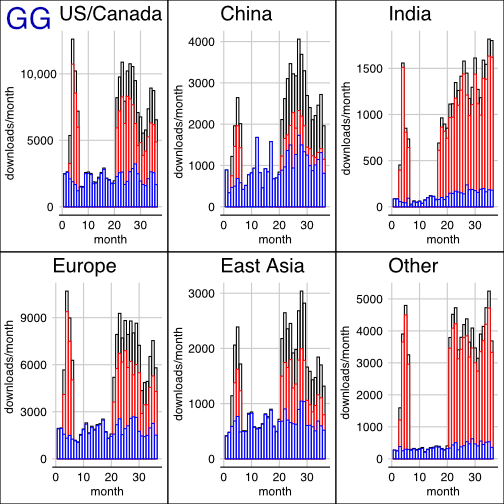} \ \ 
 \end{center}
\caption{Downloads of articles from the Annual Review of Genetics and
  Human Genomics (GG) per month for the period 2020-2022, for Inst and
  Ghost readers, for six
  different regions of the world, as described in the text.  The
  notation is as in Fig.~\ref{fig:CBglobal}.}
\label{fig:GGglobal}
\end{figure}

\begin{figure}[p]
\begin{center}
 \includegraphics[width=0.9\hsize]{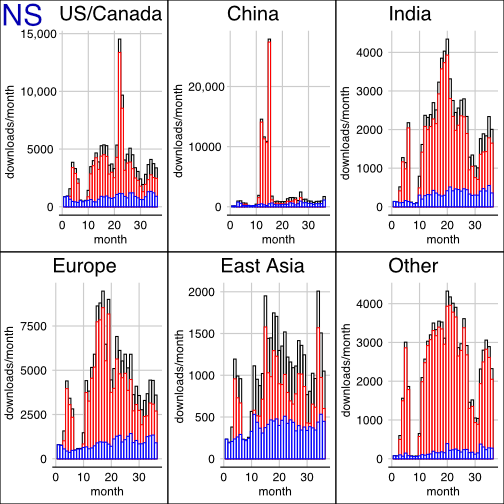} \ \ 
 \end{center}
\caption{Downloads of articles from the Annual Review of Nuclear and
  Particle Science (NS) per month for the period 2020-2022, for Inst and
  Ghost readers, for six
  different regions of the world, as described in the text.  The
  notation is as in Fig.~\ref{fig:CBglobal}.}
\label{fig:NSglobal}
\end{figure}

\begin{figure}[p]
\begin{center}
 \includegraphics[width=0.9\hsize]{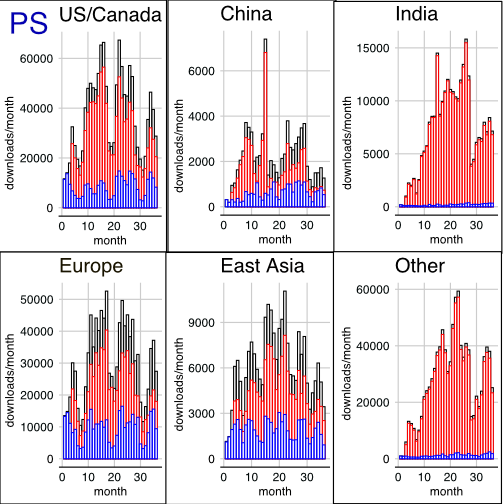} \ \ 
 \end{center}
\caption{Downloads of articles from the Annual Review of Political
  Science (PS) per month for the period 2020-2022, for Inst and
  Ghost readers, for six
  different regions of the world, as described in the text.    The
  notation is as in Fig.~\ref{fig:CBglobal}.}
\label{fig:PSglobal}
\end{figure}

\begin{figure}[p]
\begin{center}
 \includegraphics[width=0.9\hsize]{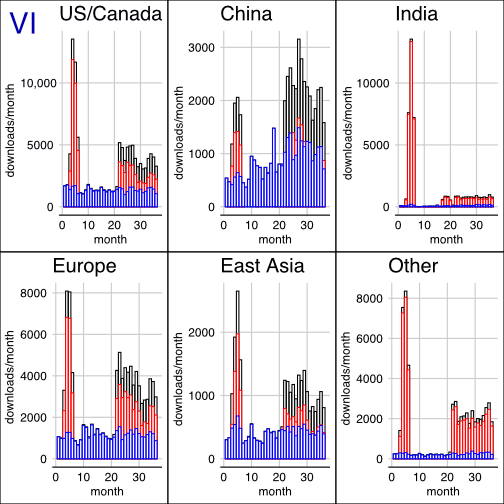} \ \ 
 \end{center}
\caption{Downloads of articles from the Annual Review of Virololgy
  (VI) per month for the period 2020-2022, for Inst and
  Ghost readers, for six
  different regions of the world, as described in the text.    The
  notation is as in Fig.~\ref{fig:CBglobal}.}
\label{fig:VIglobal}
\end{figure}

\begin{figure}
\begin{center}
 \includegraphics[width=0.4\hsize]{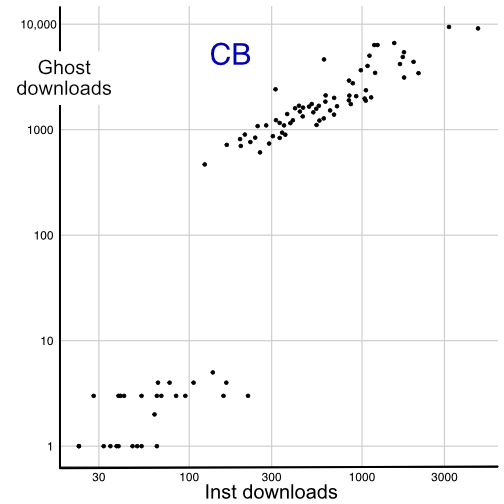} \ \ 
 \includegraphics[width=0.4\hsize]{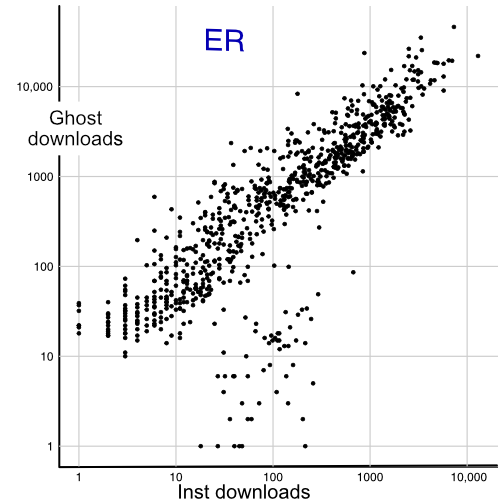} \\
 \includegraphics[width=0.4\hsize]{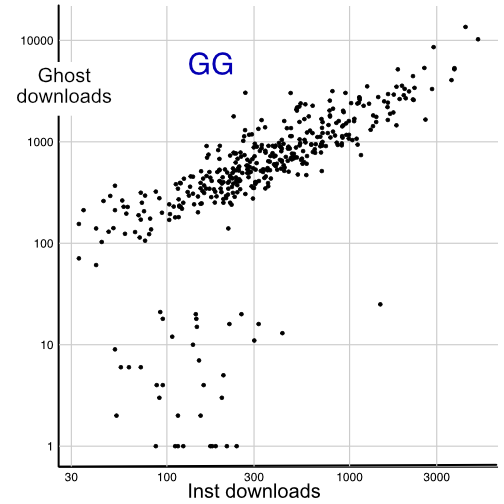} \ \
 \includegraphics[width=0.4\hsize]{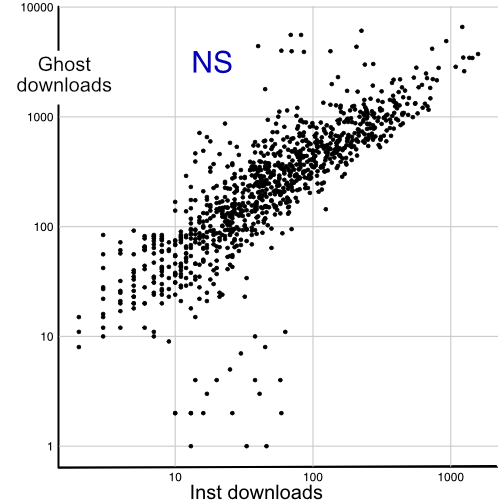}\\
 \includegraphics[width=0.4\hsize]{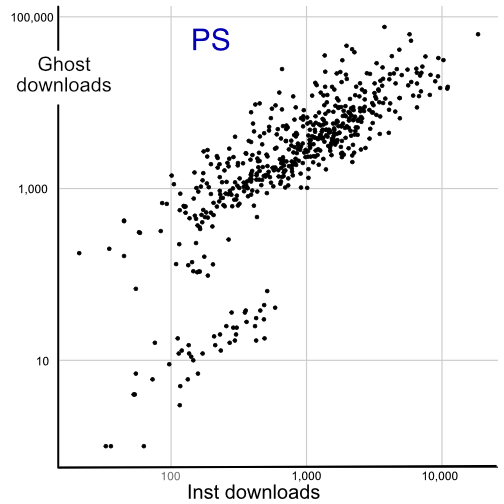} \ \
 \includegraphics[width=0.4\hsize]{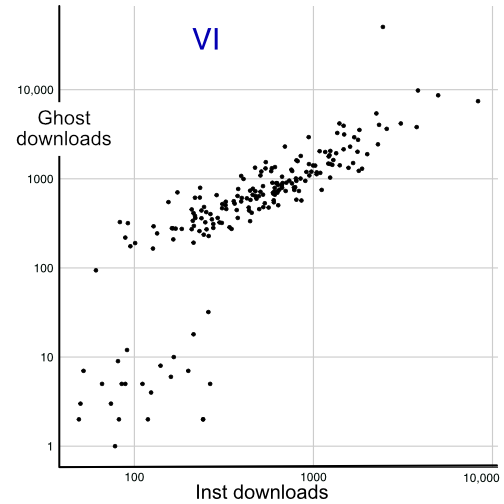}
\end{center}
\caption{Scatter plot of the number of Inst downloads vs. Ghost
  downloads for the other 6 AR journals made Open Access under S2O. In
  each plot, each point corresponds to one paper.}
\label{fig:scatters}
\end{figure}

\begin{figure}
\begin{center}
 \includegraphics[width=0.4\hsize]{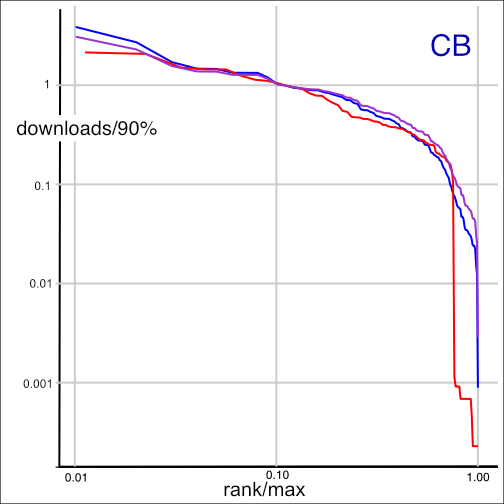} \ \ 
 \includegraphics[width=0.4\hsize]{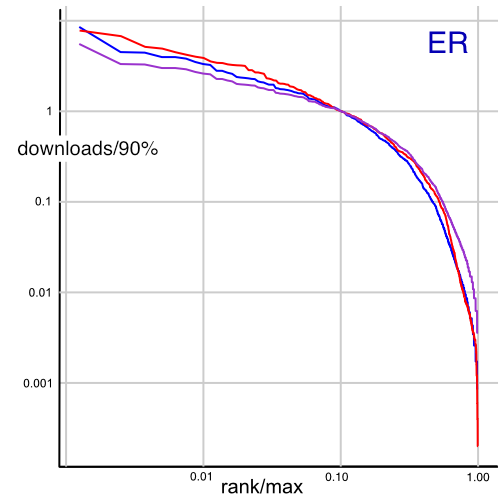} \\
 \includegraphics[width=0.4\hsize]{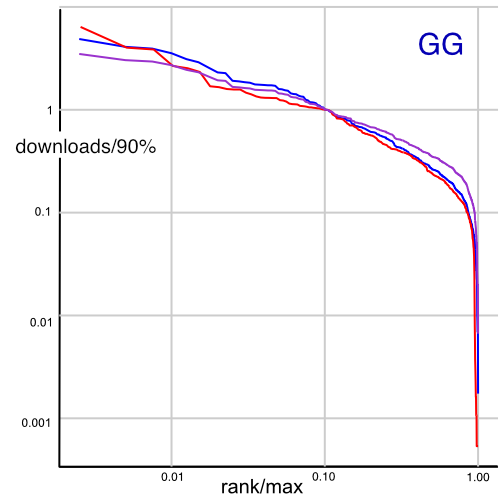} \ \
 \includegraphics[width=0.4\hsize]{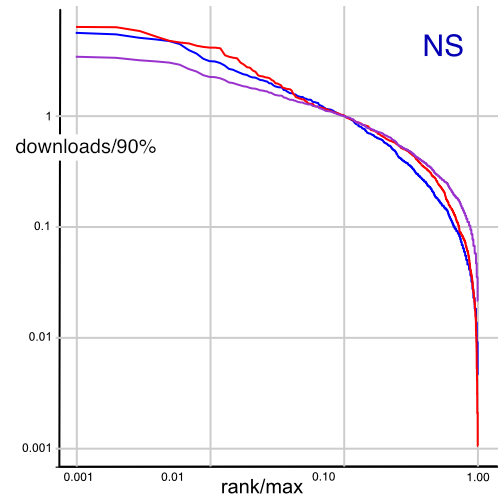}\\
 \includegraphics[width=0.4\hsize]{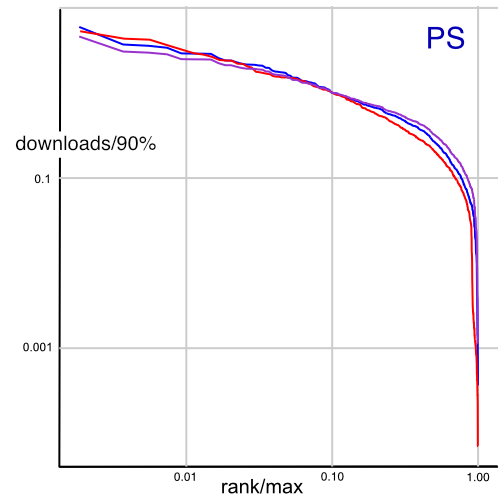} \ \
 \includegraphics[width=0.4\hsize]{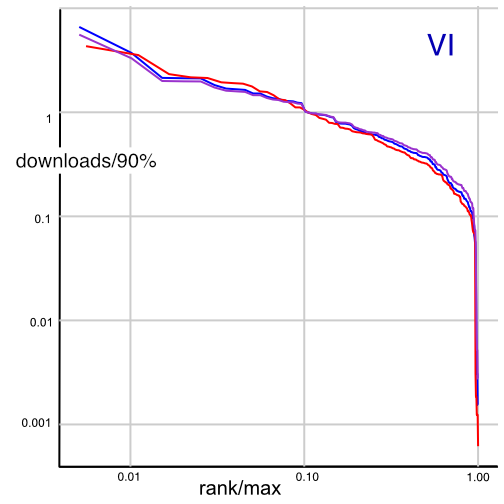}
\end{center}
\caption{Fits to the Zipf curves  for Ghost
  downloads  for the other 6 AR journals made Open Access under
  S2O. The blue and red curves are the Inst and Ghost Zipf curves,
  with the number of downloads rescaled to the number of downloads at
  the 90th percentile.  The purple curve show the transform of the
  Inst curve with the number of Inst downloads converted to Ghost
  downloads using eq. \leqn{GDID}.}
\label{fig:fits}
\end{figure}

\end{document}